\documentstyle[12pt]{article}
\begin{document}
\hfill{NCKU-HEP-97-05}\par
\hfill{hep-ph/9709236}
\vskip 0.5cm
\begin{center}
{\large {\bf A modified BFKL equation with unitarity}}
\vskip 1.0cm
Hsiang-nan Li
\vskip 0.5cm
Department of Physics, National Cheng-Kung University, \par
Tainan, Taiwan, Republic of China
\end{center}
\vskip 1.0cm

\vskip 1.0cm
%\baselineskip=2\baselineskip

\centerline{\bf Abstract}
\vskip 0.3cm

We propose a modified Balitskii-Fadin-Kuraev-Lipatov equation from the
viewpoint of the resummation technique, which satisfies the unitarity
bound. The idea is to relax the strong rapidity ordering and to restrict
phase space for real gluon emissions in the evaluation of the BFKL kernel.
It is shown that the power-law rise of the gluon distribution function
with the small Bjorken variable $x$ turns into a logarithmic rise
at $x\to 0$.
\newpage

It is known that the Balitskii-Fadin-Kuraev-Lipatov (BFKL) equation
\cite{BFKL} sums leading logarithms $\ln (1/x)$, $x$ being the Bjorken
variable, produced from the reggeon ladder diagrams with the rung gluons 
obeying the strong rapidity ordering. The gluon distribution function, 
governed by this equation, is found to increase at small $x$, which
has been confirmed by the recent HERA data of the proton
structure function $F_2$ involved in deep inelastic scattering (DIS)
\cite{H1}. However, the increase is power-like, such that $F_2$ and the
DIS cross section $\sigma_{\rm tot}$ rise as a power of $x$, {\it i.e.}, 
as a power of $s$ for $x=Q^2/s$, $Q$ being the momentum transfer and $s$ 
the total energy. This behavior does not satisfy the Froissart bound
$\sigma_{\rm tot} \le {\rm const.}\times\ln^2 s$, and violates unitarity.
Hence, the BFKL equation can not be the final theory for small $x$ physics.
Though it has been expected that the inclusions of next-to-leading 
$\ln(1/x)$ \cite{FL} and of higher-twist effects from the exchange of 
multiple ladders (pomerons) \cite{MP} may soften the BFKL rise, the 
attempts have not yet led to a concrete conclusion.
In \cite{BF} a renormalization group (RG) improved double factorization
formula was employed, and the obtained structure functions 
satisfy the unitarity bound. However, the logarithmic rise appears as
the consequence of the RG, instead of BFKL, evolution.

In this letter we shall propose a modified
BFKL equation from the viewpoint of the resummation technique \cite{CS},
which satisfies the unitarity bound. This technique was developed
originally for the organization of double logarithms $\ln^2 Q$, and
applied recently to the all-order summations of various large logarithms
contained in parton distribution functions. It has been shown that 
the known evolution equations, such as the
Dokshitzer-Gribov-Lipatov-Altarelli-Parisi (DGLAP) equation
\cite{AP}, which sums single logarithms $\ln Q$, the BFKL equation
\cite{BFKL}, and the
Ciafaloni-Catani-Fiorani-Marchesini (CCFM) equation \cite{CCFM}, which
embodies the above two equations, can be derived in a simple
and unified way \cite{L3}.

Following \cite{L3} we have presented a straightforward application of the
resummation technique \cite{L4}. Since the BFKL equation is independent of 
$Q$, its predictions are insensitive to the variation of $Q$. However, the 
HERA data for $F_2(x,Q^2)$ exhibit a stronger $Q$ dependence \cite{H1}: 
The rise at low $x$ is slower for smaller $Q$. It is plausible that the 
experiments have not explored the region with $x$ small enough to implement 
the BFKL evolution. To explain the data, the $Q$-dependent DGLAP equation 
is then combined in some way. For example, the CCFM equation was employed 
\cite{KMS}. Instead, we introduced the $Q$ dependence into the BFKL 
equation by cutting off the longitudinal component of the real gluon 
momentum at $Q$, when evaluating the BFKL kernel. The gluon distribution 
function derived from this $Q$-dependent BFKL equation leads to predictions 
of $F_2$, which are well consistent with the data. Unfortunately, the 
predictions still rise as a power of $x$ at $x\to 0$. Here we shall further 
modify the BFKL equation by relaxing the strong rapidity ordering in 
addition to restricting phase space of real gluon emissions. It is found 
that the resultant correction is negative, and the power-law rise of the 
gluon distribution function turns into a logarithmic rise at $x\to 0$.

First, we review the derivation of the BFKL equation using the resummation
technique \cite{L3,L4}. The unintegrated gluon distribution function
$F(x,k_T)$, describing the probability of a parton carrying a longitudinal
momentum fraction $x$ and transverse momenta ${\bf k}_T$, is defined by
\begin{eqnarray}
F(x,k_T)&=&\frac{1}{p^+}\int\frac{dy^-}{2\pi}\int\frac{d^2y_T}{4\pi}
e^{-i(xp^+y^--{\bf k}_T\cdot {\bf y}_T)}
\nonumber \\
& &\times\frac{1}{2}\sum_\sigma\langle p,\sigma|
F^+_\mu(y^-,y_T)F^{\mu+}(0)|p,\sigma\rangle\;,
\label{deg}
\end{eqnarray}
in the axial gauge $n\cdot A=0$, $n$ being a gauge vector with $n^2\not=0$.
The ket $|p,\sigma\rangle$ denotes the incoming proton with the light-like
momentum $p^\mu=p^+\delta^{\mu +}$ and spin $\sigma$. An average over color
is understood. $F^+_\mu$ is the field tensor.

For a fixed parton momentum $k^+$, $F$ varies with $p^+$
implicitly through $x=k^+/p^+$, and we have $p^+dF/dp^+= -xdF/dx$.
That is, the derivative of $F$ with respect to $x$ is related to the
derivative with respect to $p^+$.
Because of the scale invariance in $n$ of the gluon propagator 
$-iN^{\mu\nu}(l)/l^2$ with
\begin{equation}
N^{\mu\nu}=g^{\mu\nu}-\frac{n^\mu l^\nu+n^\nu l^\mu}
{n\cdot l}+n^2\frac{l^\mu l^\nu}{(n\cdot l)^2}\;,
\label{gp}
\end{equation}
the logarithmic corrections to $F$ must depend on the ratio
$(k\cdot n)^2/n^2\propto (p\cdot n)^2/n^2$, if assuming
$n=(n^+,n^-,{\bf 0})$. Hence, there exists a chain rule relating $p^+d/dp^+$
to $d/dn$,
\begin{eqnarray}
p^+\frac{d}{dp^+}F=-\frac{n^2}{v\cdot n}v_\beta\frac{d}{dn_\beta}F\;,
\label{cph}
\end{eqnarray}
$v_\beta=\delta_{\beta +}$ being a vector along $p$.
The operator $d/dn_\beta$ applies to a gluon propagator, giving
\begin{equation}
\frac{d}{dn_\beta}N^{\nu\nu'}=
-\frac{1}{n\cdot l}(l^\nu N^{\beta\nu'}+l^{\nu'} N^{\nu\beta})\;.
\label{dgp}
\end{equation}
The loop momentum $l^\nu$ ($l^{\nu'}$)
contracts with a vertex in $F$, which is then replaced by a special vertex 
${\hat v}_\beta=n^2v_\beta/(v\cdot nn\cdot l)$.
This special vertex can be read off the combination of
Eqs.~(\ref{cph}) and (\ref{dgp}).

The contraction of $l^\nu$ leads to the Ward identity,
\begin{equation}
l^\nu\frac{-iN^{\alpha\mu}(k+l)}{(k+l)^2}\Gamma_{\mu\nu\lambda}
\frac{-iN^{\lambda\gamma}(k)}{k^2}
=-i\left[\frac{-iN^{\alpha\gamma}(k)}{k^2}-
\frac{-iN^{\alpha\gamma}(k+l)}{(k+l)^2}\right]\;,
\label{ward}
\end{equation}
$\Gamma_{\mu\nu\lambda}$ being the triple-gluon vertex.
Summing all the diagrams with different differentiated gluons, those
embedding the special vertices cancel by pairs, leaving the one in which
the special vertex moves to the outer end of the parton line \cite{CS}.
We then obtain the derivative,
\begin{equation}
-x\frac{d}{dx}F(x,k_T)=2{\bar F}(x,k_T)\;,
\label{df}
\end{equation}
described by Fig.~1(a), where the new function $\bar F$ contains one
special vertex represented by a square. The coefficient 2 comes from the
equality of the new functions with the special vertex on either of the two
parton lines. Equation (\ref{df}) is an exact
consequence of the Ward identity without approximation \cite{CS}. An
approximation will be introduced, when ${\bar F}$ is related to $F$ by
factorizing out the subdiagram containing the special vertex, such that
Eq.~(\ref{df}) reduces to a differential equation of $F$.

It is known that factorization holds only in leading regions. The leading
regions of the loop momentum $l$ flowing through the special vertex are soft
and hard, since the factor $1/n\cdot l$ with $n^2\not=0$ in 
${\hat v}_\beta$
suppresses collinear enhancements \cite{CS}. For soft and hard $l$, the
subdiagram is factorized according to
Figs.~1(b) and 1(c), respectively. A straightforward
evaluation shows that the contribution from the first diagram of Fig.~1(c)
is less important in the considered region with small $k^+$. Therefore,
we drop Fig.~1(c), and concentrate on Fig.~1(b).
The color factor is extracted from the relation 
$f_{abc}f_{bdc}=-N_c\delta_{ad}$, where the
indices $a,b,\dots$ have been indicated in Fig.~1(b), and $N_c=3$ is the
number of colors. The factorization formula is written as
\begin{eqnarray}
{\bar F}_{\rm soft}(x,k_T)&=&
iN_cg^2\int\frac{d^{4}l}{(2\pi)^4}
\Gamma_{\mu\nu\lambda}{\hat v}_\beta
[-iN^{\nu\beta}(l)]
\frac{-iN^{\lambda\gamma}(xp)}{-2xp\cdot l}
\nonumber \\
& &\times\left[2\pi i\delta(l^2)F(x+l^+/p^+,|{\bf k}_T+{\bf l}_T|)
+\frac{\theta(k_T^2-l_T^2)}{l^2}F(x,k_T)\right],
\nonumber\\
& &
\label{kf}
\end{eqnarray}
where $iN_c$ comes from the product of the overall coefficient $-i$ in
Eq.~(\ref{ward}) and the color factor $-N_c$ extracted above.
The triple-gluon vertex for vanishing $l$ is given by
\begin{equation}
\Gamma_{\mu\nu\lambda}=
-g_{\mu\nu}xp_{\lambda}-g_{\nu\lambda}xp_{\mu}+2g_{\lambda\mu}xp_{\nu}\;.
\label{tri}
\end{equation}
The denominator $-2xp\cdot l$ is the consequence of the eikonal
approximation for the gluon propagator, $(xp-l)^2\approx -2xp\cdot l$. The
first term in the brackets corresponds to the real gluon emission, where
$F(x+l^+/p^+,|{\bf k}_T+{\bf l}_T|)$ implies that the parton coming out of
the proton carries the momentum components $xp^++l^+$ and
${\bf k}_T+{\bf l}_T$ in order to radiate a real gluon of momentum $l$. The
second term corresponds to the virtual gluon emission, where the $\theta$
function sets the upper bound of $l_T$ to $k_T$ to ensure a soft momentum
flow.

It can be shown that the contraction of $p$ with a vertex in the
quark box diagram the partons attach, or with a vertex in the gluon
distribution function, leads to a contribution down by a power $1/s$,
$s=(p+q)^2$, compared to the contribution from the contraction with
${\hat v}_\beta$. Following this observation, Eq.~(\ref{kf}) is reexpressed
as
\begin{eqnarray}
{\bar F}_{\rm soft}(x,k_T)&=&
iN_cg^2\int\frac{d^{4}l}{(2\pi)^4}N^{\nu\beta}(l)
\frac{{\hat v}_\beta v_\nu}{v\cdot l}
\left[2\pi i\delta(l^2)F(x+l^+/p^+,|{\bf k}_T+{\bf l}_T|)\right.
\nonumber \\
& &\left.+\frac{\theta(k_T^2-l_T^2)}{l^2}F(x,k_T)\right]\;.
\label{kf1}
\end{eqnarray}
The eikonal vertex $v_\nu$ comes from the last term $xp_\nu$ (divided by
$xp^+$) in Eq.~(\ref{tri}), and the eikonal propagator $1/v\cdot l$ from
$1/(xp\cdot l)$, which is represented by a double line in Fig.~1(b).
The remaining metric tensor $g^{\mu\lambda}$ has been absorbed into $F$.

To derive the conventional BFKL equation, we simply assume the strong
rapidity ordering, $x+l^+/p^+ \gg x$, for the real gluon emission, namely,
approximate $F(x+l^+/p^+,|{\bf k}_T+{\bf l}_T|)$ by its dominant value
$F(x,|{\bf k}_T+{\bf l}_T|)$. Performing the integrations over $l^-$ and
$l^+$ to infinity, and substituting
${\bar F}\approx {\bar F}_{\rm soft}$ into Eq.~(\ref{df}), we arrive at
\begin{eqnarray}
\frac{dF(x,k_T)}{d\ln(1/x)}=
{\bar \alpha}_s\int\frac{d^{2}l_T}{\pi l_T^2}
\left[F(x,|{\bf k}_T+{\bf l}_T|)-\theta(k_T^2-l_T^2)F(x,k_T)\right]\;,
\label{bfkl}
\end{eqnarray}
with ${\bar \alpha}_s=N_c\alpha_s/\pi$, which is the BFKL equation. 
To obtain the $Q$-dependent BFKL equation, we truncate $l^+$ at $Q/\sqrt{2}$
for the real gluon emission in Eq.~(\ref{kf1}), giving
\begin{eqnarray}
{\bar F}_{\rm soft}(x,k_T)
&=&\frac{{\bar\alpha}_s}{2}\int\frac{d^{2}l_T}{\pi}
\left[\int_0^{Q/\sqrt{2}} dl^+\frac{2l^+ n^2}{(2n^-l^{+2}+n^+l_T^2)^2}
\right.
\nonumber \\
& &\left.\times F(x+l^+/p^+,|{\bf k}_T+{\bf l}_T|)
-\frac{\theta(k_T^2-l_T^2)}{l_T^2}F(x,k_T)\right]\;.
\label{kf0}
\end{eqnarray}
The motivation is that the vanishing of $F(x+l^+/p^+)$ at large momentum
fraction constrains $l^+$ to go to infinity. 
Applying the strong rapidity ordering, Eq.~(\ref{kf0}) leads to
\begin{eqnarray}
\frac{dF(x,k_T)}{d\ln(1/x)}&=&
{\bar \alpha}_s\int\frac{d^{2}l_T}{\pi l_T^2}
\left[F(x,|{\bf k}_T+{\bf l}_T|)-\theta(k_T^2-l_T^2)F(x,k_T)\right]
\nonumber \\
& &-{\bar \alpha}_s\int\frac{d^{2}l_T}{\pi}
\frac{F(x,|{\bf k}_T+{\bf l}_T|)}{l_T^2+Q^2}\;,
\label{bfklq}
\end{eqnarray}
for the choice $n=(1,1,{\bf 0})$. The 
last term, corresponding to the upper bound of $l^+$, can be regarded as 
bringing in some higher-power contributions, which
moderate the BFKL rise at low $Q$.

As stated before, the $Q$-dependent BFKL equation (\ref{bfklq}), though
phenomenologically successful, predicts a power-law rise for the
gluon distribution function at $x\to 0$, which violates
unitarity. We point out that the assumption of the strong rapidity ordering
may be the cause for the unitarity violation. For most
values of $l^+$, $F(x+l^+/p^+,|{\bf k}_T+{\bf l}_T|)$ is much smaller than
$F(x,|{\bf k}_T+{\bf l}_T|)$, and replacing the former by the latter
in the whole integration range of $l^+$ overestimates the contribution
from real gluon emissions. Hence, we shall not adopt the assumption
and employ Eq.~(\ref{kf0}) directly. Using the variable change
$l^+=y\sqrt{x}p^+$, we obtain
\begin{eqnarray}
\frac{dF(x,k_T)}{d\ln(1/x)}&=&{\bar \alpha}_s\int\frac{d^{2}l_T}{\pi}
\left[\int_0^1 dy\frac{2y Q^2}{(y^2Q^2+l_T^2)^2}
F(x+y\sqrt{x},|{\bf k}_T+{\bf l}_T|)\right.
\nonumber \\
& &\left.-\frac{\theta(k_T^2-l_T^2)}{l_T^2}F(x,k_T)\right]\;,
\label{bfklu}
\end{eqnarray}
where the upper bound of $y$ is determined by the kinematic relation
$Q\approx \sqrt{2x}p^+$. Equation (\ref{bfklu}) is the modified
BFKL equation we shall investigate in more details below.

An initial condition $F(x_0,k_T)= F^{(0)}(x_0,k_T)$ must be assumed when
solving Eq.~(\ref{bfklu}), $x_0$ being the initial momentum fraction below
which $F$ evolves according to the BFKL equation.
For instance, a ``flat" gluon distribution function \cite{KMS}
\begin{equation}
F^{(0)}(x,k_T)=3\frac{N_g}{1\;{\rm GeV}^2}(1-x)^5
\exp[-k_T^2/(1\;{\rm GeV}^2)]\;,
\label{ib}
\end{equation}
with $N_g$ a normalization constant, has been proposed. Hence, the
initial function $F^{(0)}(x+y\sqrt{x},|{\bf k}_T+{\bf l}_T|)$ should be
substituted for $F(x+y\sqrt{x},|{\bf k}_T+{\bf l}_T|)$ in Eq.~(\ref{bfklu})
as $x+y\sqrt{x}> x_0$.

In order to simplify the analysis, the $\theta$ function for the
virtual gluon emission is replaced by $\theta(Q_0^2-l_T^2)$ \cite{L4},
where the parameter $Q_0$ can be determined from data fitting. This
modification is acceptable, because the virtual gluon contribution only
plays the role of a soft regulator for the real gluon emission, and setting
the cutoff of $l_T$ to $Q_0$ serves the same purpose. Reexpressing the
integrand $F(x+y\sqrt{x},|{\bf k}_T+{\bf l}_T|)$ as
\begin{eqnarray}
F(x,|{\bf k}_T+{\bf l}_T|)
+\left[F(x+y\sqrt{x},|{\bf k}_T+{\bf l}_T|)
-F(x,|{\bf k}_T+{\bf l}_T|)\right]\;,
\label{corr}
\end{eqnarray}
and Fourier transforming Eq.~(\ref{bfklu}) into the $b$ space conjugate to
$k_T$, we derive
\begin{eqnarray}
\frac{d{\tilde F}(x,b)}{d\ln(1/x)}&=&S(b,Q){\tilde F}(x,b)
\nonumber \\
& &+2{\bar \alpha}_s(1/b)Qb\int_0^1 dy K_1(yQb)
[{\tilde F}(x+y\sqrt{x},b)-{\tilde F}(x,b)]\;.
\nonumber \\
& &
\label{bfb}
\end{eqnarray}
where 
\begin{equation}
S(b,Q)=-2{\bar\alpha}_s(1/b)\left[\ln(Q_0 b)+\gamma-\ln 2+K_0(Qb)\right]
\label{es}
\end{equation}
comes from the combination of the first term in
Eq.~(\ref{corr}) and the virtual gluon emission term. $K_0$ and $K_1$ are
the Bessel functions, and $\gamma$ the Euler constant. The argument of
$\alpha_s$ has been set to the natural scale $1/b$. The second row of 
Eq.~(\ref{bfb}) is identified as
the negative correction from relaxing the strong rapidity ordering.

Before solving Eq.~(\ref{bfb}), we extract the behavior
of ${\tilde F}$ analytically.
Substituting a guess ${\tilde F}\propto x^{-\lambda}$ into Eq.~(\ref{bfb}),
$\lambda$ being a parameter, we obtain
\begin{equation}
\lambda=S+2{\bar \alpha}_sQb\int_0^1 dy K_1(yQb)
\left[\left(\frac{x+y\sqrt{x}}{x}\right)^{-\lambda}-1\right]\;.
\end{equation}
It can be numerically verified that a solution of $\lambda$, 
$0< \lambda < S$, exists for $x<x_0$. That is, ${\tilde F}$ increases as 
a power of $x$, consistent with the results from the conventional and 
$Q$-dependent BFKL equations. The correction term diverges as $x\to 0$, and 
no solution of $\lambda$ is allowed, implying that ${\tilde F}$ can not 
maintain the power-law rise at extremely small $x$. We then try another 
guess ${\tilde F}|_{x\to 0}\propto \ln(1/x)$ with a milder 
rise. It is found that the correction term, increasing as $\ln^{1.2}(1/x)$, 
only slightly dominates over the first term $S{\tilde F}\propto \ln(1/x)$ 
(the derivative term $-xd{\tilde F}/dx\propto 1$ is negligible), and 
Eq.~(\ref{bfb}) holds approximately. At last, we assume 
${\tilde F}|_{x\to 0}\propto$ const. as a test. In this trial the first term 
becomes dominant, and the correction term and the derivative term vanish,
{\it i.e.}, no const.$\not=0$ exists. These observations indicate that
${\tilde F}$ should rise as $\ln(1/x)$ at most when $x$ approaches zero. 
We conclude that the modified BFKL
equation predicts a rapid power-like rise of ${\tilde F}$ 
for $x< x_0$ and a milder logarithmic rise at $x\to 0$.

The behaviors of other quantities can be
deduced. The gluon density, defined by
\begin{eqnarray}
xg(x,Q^2)=\int_0^Q\frac{d^2 k_T}{\pi}F(x,k_T)=
2Q\int_0^{\infty} db J_1(Qb){\tilde F}(x,b)\;,
\label{gbf}
\end{eqnarray}
possesses a similar dependence on $x$, where $J_1$
is the Bessel function. The structure function $F_2$ is written, in
terms of the $k_T$-factorization theorem, as
\begin{equation}
F_2(x,Q^2)=\int_x^1 \frac{d\xi}{\xi}\int \frac{d^2k_T}{\pi}
H(x/\xi,k_T,Q)F(\xi,k_T)\;,
\label{f2}
\end{equation}
where the hard scattering subamplitude $H$ denotes the contribution from
quark box diagrams. Because of $F(\xi,k_T)\le \ln(1/\xi)$ at $\xi \to 0$,
$F_2$ rises as $\ln^2(1/x)$ at most as indicated by the $\xi$
integration, and thus satisfies the unitarity bound.

We present the explicit numerical results of the gluon density
$xg$ in Fig.~2 with $N_g\approx 3.65$ and $Q_0\approx 0.3$ \cite{L4},
which, obviously, behave in the way stated above. The deviation of the
curve from a straight line at $x\to 0$ is the consequence of relaxing
the strong rapidity ordering. The slower rise
of the curve for a smaller $Q$ is due to the truncation of $l^+$
at $Q/\sqrt{2}$. The comparision of our predictions for $F_2$ with
the experimental data \cite{H1} and the connection of the modified BFKL
equation (\ref{bfklu}) to the other approaches in the literature will be
published elsewhere.

This work was supported by the National Science Council of R.O.C. under the 
Grant No. NSC-87-2112-M-006-018.

\newpage
\centerline{\large \bf Figure Caption}
\vskip 0.5cm

\noindent
{\bf FIG. 1.} (a) The derivative $-xdF/dx$ in the axial gauge.
(b) The soft structure and (c) the ultraviolet structure of the
$O(\alpha_s)$ subdiagram containing the special vertex.
\vskip 0.5cm

\noindent
{\bf FIG. 2.} The dependence of $xg$ on $x$ for
$Q^2=8.5$ GeV$^2$ (dotted line), 15 GeV$^2$ (dashed line), and 20 GeV$^2$
(solid line).

\end{document}